\documentclass{ifacconf}

\usepackage{graphicx}      
\usepackage[round]{natbib}        
\usepackage{amssymb,amsmath}

\usepackage{color}
\usepackage{enumerate}

\begin{document}

\begin{frontmatter}

\title{Assessing Linear Control Strategies for Zero-Speed Fin Roll Damping}


\author[Nv,Leti]{Nikita Savin}
\author[Nv,Leti]{Elena Ambrosovskaya}
\author[Nv]{Dmitry Romaev}
\author[Po]{Anton Proskurnikov}

\address[Nv]{Navis JSC, St. Petersburg, Russia}
\address[Leti]{St. Petersburg Electrotechnical University ``LETI'', Russia}
\address[Po]{Politecnico di Torino, Italy}
\thanks[footnoteinfo]{
Email: {\tt\small \{n.savin,e.ambrosovskaya,d.romaev\}@navis.spb.ru, anton.p.1982@ieee.org}
}

\begin{abstract}
Roll stabilization is a critical aspect of ship motion control, particularly for vessels operating in low-speed or zero-speed conditions, where traditional hydrodynamic fins lose their effectiveness. In this paper, we consider a roll damping system, developed by Navis JSC, based on two actively controlled zero-speed  fins. Unlike conventional fin stabilizers, zero-speed fins employ a drag-based mechanism and active oscillations to generate stabilizing forces even when the vessel is stationary. We propose a simple linear control architecture that, however, accounts for nonlinear drag forces and actuator limitations. Simulation results on a high-fidelity vessel model used for HIL testing demonstrate the effectiveness of the proposed approach.
\end{abstract}

\end{frontmatter}

\section{Introduction}

Roll damping is a long‑studied problem in ship motion control -- essential on passenger vessels, where comfort is paramount, and equally critical on commercial and specialized ships, where excessive roll can compromise cargo safety, reduce operational efficiency,  and hinder precision operations such as offshore loading, seismic surveying, or ROV deployment. A wide spectrum of anti‑roll methods has been developed, from passive bilge keels and tuned anti‑roll tanks to active stabilizers; see~\citep{PerezBlanke2012129,perez2006ship} for a comprehensive review.

One approach to roll damping -- often called \emph{rudder roll damping}~\citep{VANAMERONGEN1990679} -- exploits the ship’s vectoring actuators, such as conventional rudders, azimuthing thrusters, and Voith‑Schneider propellers (VSPs), to generate the anti‑roll moments needed for stabilization. This approach eliminates the need for dedicated roll‑stabilizing hardware.
 However, because the same actuators also serve heading and position control, it requires a careful trade-off between steering performance and roll-damping effectiveness~\citep{perez2006ship,Kapitanyuk2020}. Moreover, the nonminimum-phase characteristics of roll dynamics impose fundamental limits on achievable closed‑loop performance~\citep{GoodwinPerez914671}. Since thrusters and rudders are primarily optimized for steady course‑keeping and station‑keeping, commanding them to produce high‑frequency lateral forces at a vessel’s natural roll frequency significantly accelerates mechanical wear and reduces their service life. 

Therefore, dedicated roll stabilizers are generally favored over rudder‑based roll damping. These solutions span from passive and semi‑passive devices -- bilge keels, anti‑roll tanks, and fixed or variable‑geometry fins -- to fully active systems, such as hydraulic or electric fin stabilizers and gyroscopic units, which leverage sensors and advanced control algorithms for optimal roll reduction at the expense of greater energy consumption and mechanical complexity.

For yachts and small passenger vessels, the choice is typically between a gyroscopic stabilizer and hull‑mounted stabilizer fins. 
At cruising speeds and in severe roll conditions, fin stabilizers typically outperform gyroscopic stabilizers, as fins do not face mechanical saturation limits like gyros, which can reach their maximum tilt angle. Moreover, fins require less power per unit of damping torque under these conditions. Additionally, hull-mounted fin actuators preserve valuable internal space, and generally have simpler, less demanding maintenance requirements compared to gyrostabilizers. However, traditional fins lose their effectiveness at low speeds, since the hydrodynamic lifting forces they generate decrease proportionally to the square of the vessel’s speed through the water~\citep{Kula2014}. 

Zero‑speed fins overcome this limitation by employing a drag‑based mechanism, actively oscillating (flapping) the fin to ``pump'' water and thus generate substantial stabilizing forces, even when the vessel is stationary or moving slowly. These active oscillations are inspired by the motions of fish fins and bird wings. In recent years, zero‑speed (or flapping) fin stabilizers have become increasingly popular. A growing body of research has focused on accurately modeling the hydrodynamic forces acting on fins during low‑speed operation, optimizing fin geometry, and developing advanced control algorithms to enhance performance~\citep{Liang2018simulation,Zhang2019,Song2019,Song2020,WeiYang2021,Xu2025_OE}.

In this paper, we present control and identification techniques for a ship roll-stabilization system featuring two zero-speed fins, currently under development by Navis JSC in collaboration with shipbuilding plants. The key features of the proposed algorithms are: first, the explicit consideration of the nonlinear drag force acting on the fins; and second, the enforcement of constraints on both fin angle and angular rate, as dictated by the limitations of the hydraulic actuator. Control designs that explicitly account for these nonlinearities result in complex optimization problems with non-convex constraints. These are typically addressed using advanced techniques, including robust MPC frameworks~\citep{Malekizade2016} or heuristic methods such as genetic algorithms~\citep{Song2020GA}. 

Since the roll-damping algorithm must be implemented on a low-cost microcontroller and easily adapted to changing sea conditions, it is highly desirable to keep the design as simple as possible and remain within the class of linear control strategies.
In this paper, we examine a class of linear controllers and prove the robust stability of the resulting closed-loop system in the presence of bounded disturbances using the circle criterion for incremental stability (strong contraction)~\citep{AltoCorless2013,FB:24-CTDS}. The need for an absolute stability criterion arises from the nonlinear rate–moment characteristics and the saturation of the fin angular rate.
While remaining simple, the proposed controller can still be tuned to accommodate varying sea conditions, maintaining a balance between the requirements of global stability and effective roll damping.

The remainder of the paper is organized as follows. Section~\ref{sec.model} introduces the mathematical model of the vessel's roll motion. Section~\ref{sec.contr} addresses the control design and related stability considerations. Section~\ref{sec.simul} presents simulation results that demonstrate the effectiveness of the proposed controller on a realistic vessel model used by Navis JSC for Hardware-in-the-Loop testing.

\section{Mathematical Modelling}\label{sec.model}

The ship’s roll motion is driven by the disturbance moment $M_w(t)$, primarily wave-induced, and is counteracted by the control moment $M_u(t)$, which must be delivered by actuators of sufficient speed and power.
\begin{figure}[t]
    \centering
    \includegraphics[width=0.95\linewidth]{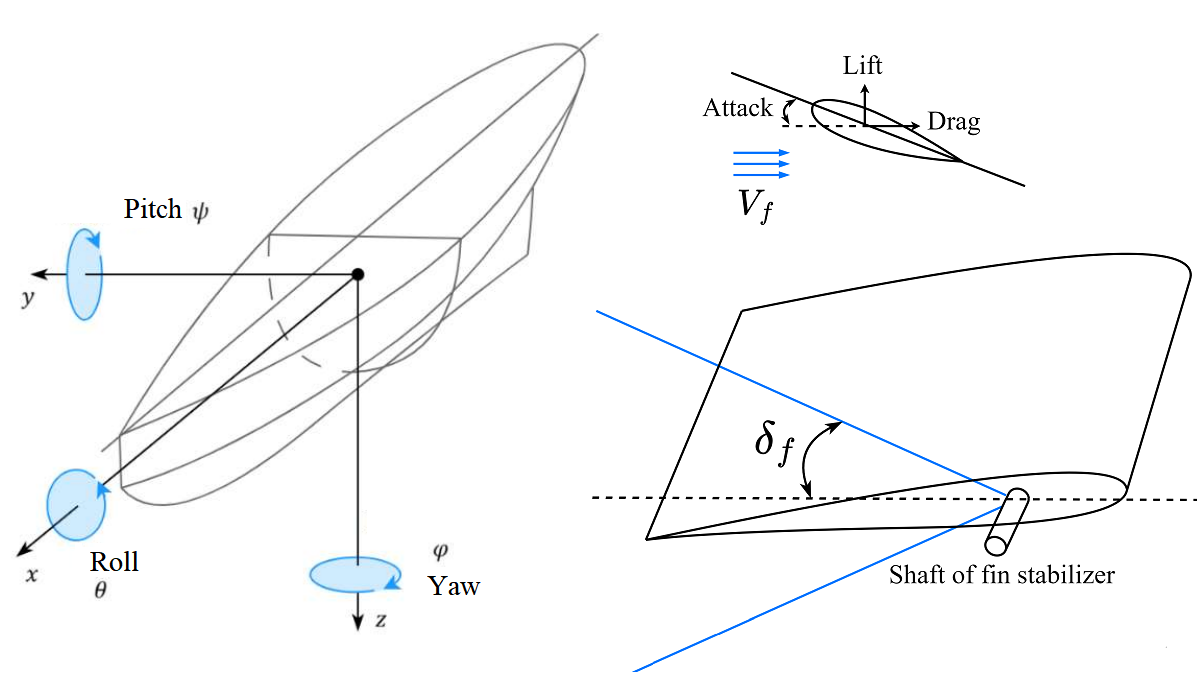}
    \caption{The ship coordinate system, fin stabilizer angle $\delta_f$}
    \label{fig:coord}
\end{figure}


\subsection{Vessel Roll Dynamics} In the standard right-handed coordinate system (Fig.~\ref{fig:coord}), the ship’s roll dynamics are
\begin{equation}
    J_{xx} \ddot \theta + f_d(\dot\theta) + mgl(\theta) = M_w(t) +M_u(t),
\end{equation}
Here, $J_{xx}$ denotes the roll moment of inertia, including added-mass contributions; and $mgl(\theta)$ is the restoring moment. The nonlinear roll-damping moment $f_d(\dot\theta)$ is often modeled parametrically~\citep{Song2019} as
\begin{equation}\label{eq.f_d}
  f_d(\dot\theta) = 2 N_1 \dot \theta + N_2 |\dot \theta| \dot \theta,
\end{equation}
where the hydrodynamic coefficients $N_1$ and $N_2$ depend on the vessel speed $V$ and typically increase as $V$ grows.
In the control‐design phase, both the righting arm $l(\theta)$ and the damping moment $f_d(\dot\theta)$ are linearized about the equilibrium point, as described below.

For anti-roll controller design, the nonlinear ship-roll and fin-actuation dynamics are typically linearized and expressed as the time-invariant model derived as follows:
\begin{enumerate}[(a)]
\item for sufficiently small roll angles, the righting arm can be approximated by  
$l(\theta)\approx h_{\theta}\,\theta$, where $h_{\theta}$ is the transverse metacentric height;
\item by applying the describing‐function method~\cite{Song2019,Chechurin2017} to the nonlinear damping moment~\eqref{eq.f_d},
it can be approximated by the linear function $f_d(\dot\theta)\approx N_{\theta}\,\dot\theta$, where the equivalent coefficient $N_{\theta}$ depends on the vessel's speed and the roll-motion amplitude and frequency;
\end{enumerate}

Under these simplifications, the roll‐motion dynamics reduce to the linear damped‐pendulum equation driven by two external inputs, namely the disturbance moment $M_w(t)$ and the control moment $M_u(t)=M_u(\delta_f(t),\dot\delta_f(t))$:
\begin{equation}\label{eq:linear}
     \ddot \theta + 2 \nu_\theta \omega_0 \dot \theta + \omega_0^2\theta = \frac{1}{J_{xx}}(M_u(t) +M_w(t)).
\end{equation}
Here, the natural roll frequency and the (dimensionless) damping coefficient are defined by
\[
\omega_0^2 = \frac{m g\,h_{\theta}}{J_{xx}}, 
\quad
\nu_{\theta} = \frac{N_{\theta}}{J_{xx}\omega_0}.
\]
The roll natural period is $T_0 = 2\pi/\omega_0$. Recall that the control moment has been found in~\eqref{eq:Mu}.



\subsection{Zero-speed Fin Stabilizers.} For fin stabilizers, the control moment $M_u(t)$ decomposes into two components: a lift‐based term that depends on inflow velocity $V_f$ and fin angle $\delta_f$, and a zero‐speed drag term that depends solely on the fin’s angular rate $\omega_f = \dot\delta_f$ \citep{Zhang2019}. Consequently, the total control moment is expressed as
\begin{equation}\label{eq:Mu}
\begin{aligned}
M_u(t) &= M_{\rm lift}(t) + M_{\rm zero}(t)\\
M_{\rm lift}&={\rho V_f^2 \over 2}  S_f l_f C_f(\delta_f),\\ 
M_{\rm zero}&=l_f S_f C_{f0} (\dot\delta_f)
\end{aligned}
\end{equation}
Here $\rho$ is the fluid density, $S_f$ is the fin area, $l_f$ the lever arm from the roll axis to the fin’s center of pressure,
$C_f$ is the
lift coefficient depending on fin profile ~\citep{PerezBlanke2012129}
and $C_{f0}$ is the zero-speed force coefficient that are found approximately as
\[
\begin{gathered}
C_f(\delta_f)\approx C^{\delta}\delta_f,\quad C^{\delta}>0,\\
C_{f0}(\dot\delta_f)\approx k\dot\delta_f|\dot\delta_f|,\quad k>0.
\end{gathered}
\]
where $C^{\delta}$ and $k$ are positive constants determined experimentally. In practice, both the fin angle and its angular rate are constrained. Consequently, the zero‐speed coefficient $C_{f0}(\dot\delta_f)$ is replaced by a more realistic nonlinear characteristic $h(\dot\delta_f)$, which will be introduced later. 

\subsubsection*{Actuation Dynamics of Fin Stabilizers.} 
\begin{figure}[htb]
    \centering
    \includegraphics[width=0.99\linewidth]{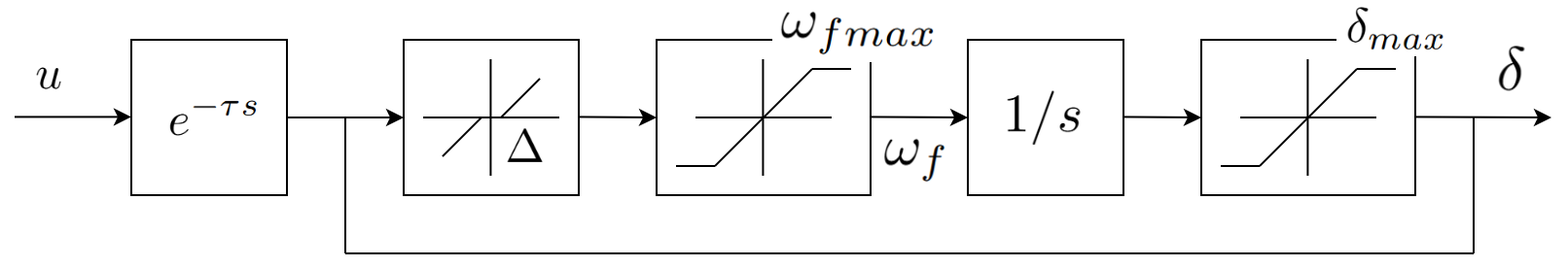}
    \caption{Fin stabilizer actuator with FFU control. }
    \label{fig:act}
\end{figure}

Fin stabilizers typically employ electrohydraulic actuators equipped with full-follow-up control (FFU), which may be implemented either within the actuator’s inner-loop controller or as an integral part of the vessel’s roll-control system.
Typically, the two symmetric fins are actuated synchronously but in anti-phase (i.e., when one deflects upward, the other deflects downward). Denoting the commanded fin angle by $u(t)$, the fin-stabilizer actuation sctructural scheme is illustrated in Fig.~\ref{fig:act}.


As shown in Fig.~\ref{fig:act}, the fin’s electrohydraulic actuator obeys the nonlinear dynamical system 
\begin{equation}
   \omega_f= \dot \delta_f = f(u(t-\tau), \delta_f),
\end{equation}
which is characterized by:  
\begin{enumerate}
    \item transport delay, $\tau$;   
    \item the servo-valve deadband width $\Delta$;
    \item maximum fin angular rate, $\omega_{f\, \max}$; 
    \item maximum fin deflection angle, $\delta_{f\, \max}$;
    \item full‐stroke travel time (minimum to maximum angle). 
\end{enumerate}
This nonlinear actuator model is implemented in the vessel-control digital twin (Section IV) and was employed, in particular, for the HIL testing of the anti-roll controller introduced in the next section.

For the control design, however, the fin’s inner‐loop actuator dynamics is often approximated by the first‐order differential equation~\citep{Malekizade2016}  
\begin{equation}\label{eq:privod-simple}
\dot\delta_f(t)=\frac{u(t)-\delta_f(t)}{T_{\delta}},
\end{equation}
where the time constant $T_{\delta}>0$ captures the combined effects of transport delay and dead‐zone nonlinearities.



\subsection{Wave disturbance}

The sea-surface elevation is modeled as a stochastic process with spectral density $S(\omega)$, which describes the distribution of wind-driven wave energy across frequencies; the Pierson–Moskowitz and JONSWAP bell-shape spectra are the most widely used models. Realizations of this process can be approximated arbitrarily closely by a polyharmonic signal, where the number of harmonics $N$ is large enough, the amplitudes and frequencies are chosen to replicate $S(\omega)$, and random phases are uniformly distributed over $[0,2\pi]$~\citep[Section 2.10]{perez2006ship}.

In view of this, we model the wave disturbance moment $M_w(t)$ as a polyharmonic signal:
\[
M_w(t) = mgh_\theta \theta_e(t),\quad
\theta_e(t) =  \sum_{i=1}^N  b_i \sin(\omega_i t  +\varphi_i)
\]
For each harmonic component $i$, $b_i$ denotes the amplitude of the effective sea–slope angle; $\omega_i=2\pi/T_i$ the angular frequency (with $T_i$ the period); 
and $\varphi_i$ the phase. 

\section{Controller design}\label{sec.contr}

Roll stabilization using conventional fins typically relies on linear control methods -- from PID and loop‐shaping to advanced $H_{\infty}$ controllers -- enabled by the ``quasi‐linear'' structure of system~\eqref{eq:linear}, in which the damping moment is linearized and the nonlinear drag‐induced moment is often neglected because it is overshadowed by the lift‐induced force (see Fig.~\ref{fig:mom} in Section~\ref{subsec.ship}).
This linearity also enables efficient optimization-based techniques, taking into account the actuator constraints and mitigating dynamic stall~\citep{Perez2008,Jimoh2021} and invariant‐set techniques~\citep{Ghaemi2009}. 

In low‐speed operation mode, roll‐damping control must both exploit the dominant nonlinear drag moment and comply with actuator limits -- namely, the maximum fin angle and slew‐rate constraints imposed by the servo‐valve dynamics. Note that the control is indirect: although the zero-speed drag moment depends on the fin’s angular rate $\omega_f = \dot\delta_f$, the commanded input to the steering system is the fin angle $\delta_f$.
This renders optimization‐based approaches challenging, as they result in \emph{non-convex} programs.
Existing approaches include robust model‐predictive control, which necessitates real‐time semidefinite programming solutions~\citep{Malekizade2016}, genetic algorithms for heuristic optimization~\citep{Song2020GA}, and hybrid schemes that integrate fuzzy and sliding‐mode control~\citep{Su2018}.

\subsection{The Proposed Linear Control Design}

Aiming for implementation on low-capacity microcontrollers, we consider the following simple linear controller:
\begin{equation}\label{eq:controller} 
u(t) = (1 - T_{\delta} c)\delta_f(t) - T_{\delta} (k_p \theta(t) + k_d \dot\theta(t)), 
\end{equation}
where $T_{\delta}$ is the time constant from the (approximate) actuator dynamics~\eqref{eq:privod-simple}, 
$c > 0$ and $k_p, k_d $ 
are tunable gains. As will be discussed, the key requirement for these coefficients is that they satisfy the circle criterion, which guarantees incremental stability (i.e., strong contraction) of the closed-loop system -- ensuring, in particular, the stability of every forced response to a bounded disturbance.

The intuition behind the controller choice is as follows. By simplifying the fin actuator dynamics to~\eqref{eq:privod-simple}, the fin angle dynamics reduce to:
\begin{equation}\label{eq:controller1} 
\dot\delta_f(t)+c\delta_f(t)=\omega_f^*(t):=-k_p \theta(t)-k_d \dot\theta(t)
\end{equation}
In a sense, $\omega_f^*$ can be interpreted as the desired fin angular rate. If $c = 0$ and neither the fin angle $\delta_f$ nor its rate $\dot\delta_f$ is saturated, then $\dot\delta_f = \omega_f^*$. The coefficient $c > 0$ introduces additional damping relative to the fin angle, making the dynamics of $\delta_f$ stable, enhancing robustness to disturbances and helping to prevent fin angle saturation.


\subsection{Accounting for System Nonlinearities}

While the control design is linear, the actual dynamics of the vessel–fin–actuator system are highly nonlinear due to the presence of several distinct nonlinearities. The first and most significant nonlinearity arises in the control moment. Assuming the ship operates at low speed—so that the lift component $M_{\rm lift}$ can be neglected—the control-related term on the right-hand side of~\eqref{eq:linear} reduces to
\begin{equation}\label{eq:m_u_k02} 
m_u(t) := \frac{1}{J_{xx}} M_u(t)\approx k_{02} \omega_f(t) |\omega_f(t)|,
\end{equation}
where $k_{02} > 0$ is constant. Additional nonlinearities arise from the dynamics of the fin's actuator (Fig.~\ref{fig:act}), which impose limits on the fin angle and angular rate.

Since analyzing systems with multiple nonlinearities is challenging, we adopt a further simplification that aligns with standard engineering practice. Specifically, we note that the moment in~\eqref{eq:m_u_k02} can only be generated within the range $\omega_f \in [-\omega_{f,\max}, \omega_{f,\max}]$, and becomes saturated at its maximal (or minimal) value
\[
h_{\max}: = k_{02} \omega_{f\,\max}^2
\]
(respectively, $-h_{\max}$) outside this interval. We denote the actual saturated moment by $h(\omega_f)$; see Fig.~\ref{fig:nlin}.
As will be shown below, the exact shape of the rate-torque curve is unimportant; the only essential properties are its boundedness and the \emph{slope restriction} as follows:
\begin{equation}\label{eq:slope}
0\leq\frac{h(\omega_1)-h(\omega_2)}{\omega_1-\omega_2}\leq k_f:=2k_{02}\omega_{f \max}=\frac{h_{\max}}{\omega_{f \max}}.
\end{equation}
\begin{figure}[htb]
    \centering
    \includegraphics[width=0.6\linewidth]{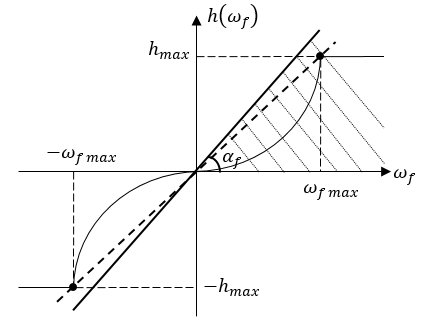}
    \caption{ Saturated zero-speed fins nonlinear rate-moment curve}
    \label{fig:nlin}
\end{figure}

The introduced saturated control torque ensures that $\theta$ and $\dot\theta$ remain bounded in view of~\eqref{eq:linear} for any bounded disturbance $M_w(t)$. In practice, this means that by choosing the gains $k_p$, $k_d$ sufficiently small and the damping coefficient $c > 0$, the fin angle $\delta_f$ and its rate $\dot\delta_f$ will not reach saturation under the proposed controller~\eqref{eq:controller}, since the dynamics of $\delta_f$ is given by~\eqref{eq:controller1}.

Under these simplifying assumptions, the dynamics of the closed-loop system is as follows
\begin{equation}\label{eq:lurie0}
\begin{gathered}
\ddot \theta + 2 n_\theta \dot\theta + \omega_0^2 \theta = h(\dot\delta_f) + m_w(t),\\
\dot\delta_f(t)+c\delta_f(t)=-k_p \theta(t)-k_d \dot\theta(t).
\end{gathered}
\end{equation}
where $h(\omega_f)$ is the saturated drag-induced control moment, and $m_w(t)$ is the external disturbance moment (both moments are scaled by $1/J_{xx}$)

\begin{figure}
    \centering
    \includegraphics[width=0.98\linewidth]{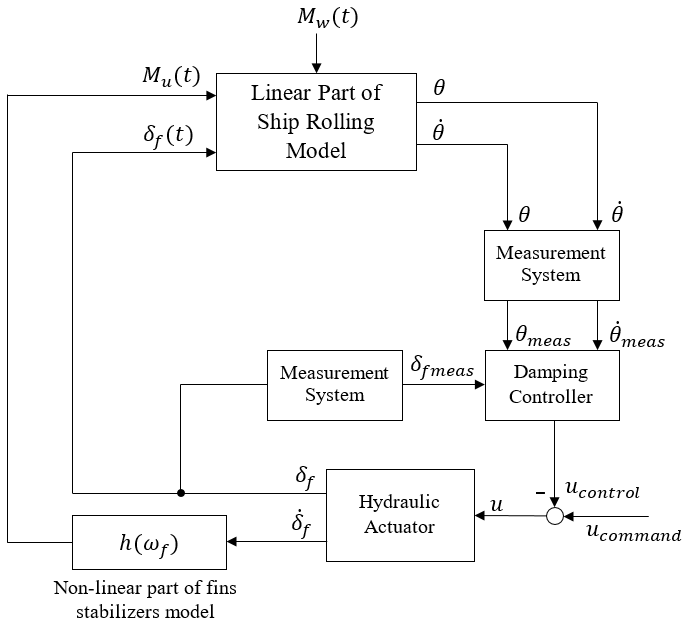}
    \caption{Plant in Lurie form}
    \label{fig:plant}
\end{figure}

The system~\eqref{eq:lurie} is the Lur'e system with one scalar nonlinearity $h(\cdot)$, satisfying the slope constraint~\eqref{eq:slope} and the external disturbance $m_w(t)$. Figure~\ref{fig:plant} illustrates the structure of the closed-loop system.

\subsection{Incremental Stability}

Note that although all solutions remain bounded under a bounded disturbance $m_w(t)$, this form of stability is relatively weak. A far more meaningful property for systems subject to disturbances is what was originally introduced by B.~P.~Demidovich as ``convergent dynamics''~\citep{Pavlov2004}, and is known in modern literature as (exponential) \emph{incremental stability}~\citep{AltoCorless2013} or strong (infinitesimal) contraction~\citep{FB:24-CTDS}.

Without introducing the general definition, we simply note that for Lur’e-type systems
\begin{equation}\label{eq:lurie}
\dot x(t)=Ax(t)+B_hh(y(t))+B_w w(t),\quad y(t)=Cx(t),    
\end{equation}
with external input $w(t)$ and nonlinear function $h(y)$, 
this property is typically established using quadratic \emph{incremental} quadratic Lyapunov functions~\citep{Pavlov2004}
$V(\Delta x)=(\Delta x)^{\top}P\Delta x$. Here, $\Delta x(t) = x(t) - {\tilde x}(t)$ denotes the discrepancy between two solutions corresponding to the same external disturbance $w(t)$, and $P = P^{\top} > 0$ is a positive definite matrix. Under certain conditions, there exists a matrix $P$ -- dependent only on the linear part of the system, but not on the specific nonlinearity within the sector or the disturbance $w(t)$ -- such that
\begin{equation}\label{eq:increment}
\frac{d}{dt}V(\Delta x(t))\leq -\varepsilon V(\Delta x(t)),
\end{equation}
meaning that the difference between any two solutions decays exponentially.

We confine ourselves to the case where $y$ is scalar and $h$ satisfies the slope restriction~\eqref{eq:slope}.
Note that for a \emph{linear} function $h(y) = ky$, the above condition reduces to the standard notion of stability. Therefore, the criterion for incremental stability can be viewed as a stronger form of the classical Nyquist criterion, guaranteeing stability for all linear gains $k \in [0, k_f]$.

Denoting the transfer function from $h$ to $y$ by
\[
W(s)=C(sI-A)^{-1}B,
\]
the following criterion for the incremental stability holds, see~\citep{AltoCorless2013}.
\begin{lem}\label{lem.1}
Assume that $A$ is a Hurwitz\footnote{A matrix is Hurwitz if all its eigenvalues have negative real parts.} matrix and that the following frequency-domain condition holds
\[
\mathop{Re} W(j\omega)<\frac{1}{k_f}\quad\forall \omega\in\mathbb{R}.
\]
Then, the Lur’e system~\eqref{eq:lurie} is incrementally stable for any nonlinear function $h$ satisfying the slope condition~\eqref{eq:slope}. Moreover, there exists a quadratic incremental Lyapunov function $V(\Delta x) = (\Delta x)^{\top} P \Delta x$ that satisfies~\eqref{eq:increment} for every two solutions $x(t),\tilde x(t)$, corresponding to the same disturbance $w(t)$. Here both $P=P^{\top}>0$ and $\varepsilon$ depend only on $(A, B, C)$ and $k_f$, but not on $h(\cdot)$ or $w(t)$.
\end{lem}

The incremental stability is important, because it guarantees a number of important properties~\citep{FB:24-CTDS}:
\begin{itemize}
\item In the absence of disturbances ($w(t) = 0$), the system admits a unique and exponentially stable equilibrium. If $h(0) = 0$, then the origin $x = 0$ is this equilibrium.
\item If $w(t)$ is periodic, then all solutions $x(t)$ converge exponentially fast to a unique forced periodic solution of the same frequency.\footnote{This property is often referred to as \emph{entrainment} and was first established by~\cite{VAY:64}.}
\item The solution $x(t)$ depends continuously on the disturbance $w(t)$: there exist $C_1, C_2 > 0$ such that  
\[
\|\Delta x(t)\| \leq C_1 e^{-\varepsilon t} \|\Delta x(0)\| + C_2 \sup_{s \in [0, t]} \|\Delta w(s)\|
\]  
for any two state trajectories $x(t)$ and $\tilde{x}(t)$ corresponding to disturbances $w(t)$ and $\tilde{w}(t)$, respectively. As usual,
$\Delta x=x-\tilde x$ and $\Delta w=w-\tilde w$.
\item In particular, if $h(0)=0$, then applying the previous inequality to $\tilde x=0$ and $\tilde w=0$, one proves that
\[
\|x(t)\|\leq C_1 e^{-\varepsilon t} \|x(0)\| + C_2 \sup_{s \in [0, t]} \|w(s)\|.
\]
\end{itemize}
Constants $C_1,C_2$ can be found explicitly and depend on $P,\varepsilon$~\citep[Theorem~3.16]{FB:24-CTDS}.

By noticing that the transfer function from $h$ to $\omega_f=\dot\delta_f$ for the system~\eqref{eq:lurie0} is found as
\[
W(s)=-\frac{s}{s+c}\frac{k_ds+k_p}{s^2 + 2 n_\theta s + \omega_0^2},
\]
as a corollary of Lemma~\ref{lem.1}, we obtain the following.
\begin{cor}
The system~\eqref{eq:lurie0} is incrementally stable if $h$ satisfies the slope constraint~\eqref{eq:slope} and
\begin{equation}\label{eq:freq}
\begin{gathered}
\mathop{Re}\tilde W(j\omega)+\frac{1}{k_f}>0,\\
\tilde W(j\omega):=-W(j\omega)=\frac{j\omega(k_dj\omega+k_p)}{(j\omega+c)\left(2 n_\theta j\omega + \omega_0^2-\omega^2\right)}.
\end{gathered}
\end{equation}
\end{cor}

It should be noted that the frequency-domain conditions for incremental stability can be replaced by linear matrix inequality (LMI) conditions~\citep{AltoCorless2013,FB:24-CTDS}, which are more convenient when the coefficients of the Lur’e system are fixed. However, when the control gains are treated as variables, the LMIs become nonconvex bilinear constraints. Unlike them, if $c > 0$ is fixed, the left-hand side of~\eqref{eq:freq} is linear in $(k_d, k_p)$, and the set of gain pairs satisfying~\eqref{eq:freq} remains convex.

\subsection{Remarks on The Gain Tuning}

Note first that the frequency-domain condition can always be satisfied by choosing the gains $k_p$ and $k_d$ sufficiently small. This is expected—if the control input is weak, it cannot destabilize an already asymptotically stable system. However, a weak control action is also insufficient for effectively damping roll oscillations. Hence, a principled approach to selecting the control gains requires solving an optimization problem that balances nonlinear stability with roll damping performance -- an objective that lies beyond the scope of this work and is the subject of ongoing research. At the same time, experiments with a realistic vessel model under various sea conditions (see next section) demonstrate that it is possible to tune the control gains such that:
(a) satisfactory damping is achieved for wave disturbances approximated by a polyharmonic process;
(b) saturation of the fin angle and fin angular rate is avoided; and
(c) the sufficient condition for incremental stability is satisfied.
As a first step, we linearize the nonlinear function $h(\omega)$ using harmonic linearization: we subject the system to a sinusoidal input of given amplitude and period (obtained via slow averaging of the disturbance), compute the equivalent gain, and then select the gains to place the poles of the resulting linearized system at the desired locations. Alternatively, one may employ more advanced control designs, such as LQR, LQG, or $H_\infty$ control. If~\eqref{eq:freq} is violated, we proportionally reduce the gains $k_p$ and $k_d$ until the Nyquist plot of $\tilde W(j\omega)$ lies entirely to the right of the vertical line $\operatorname{Re} = -\tfrac{1}{k_f}$. The coefficient $c>0$ has little impact on the frequency-domain condition~\eqref{eq:freq}, yet it helps prevent fin-angle saturation. However, $c$ must remain small; otherwise, the fin's response to changing disturbances becomes excessively sluggish. 
\color{black}


\section{Experiments}\label{sec.simul}

\begin{figure}[htb]
    \centering
    \includegraphics[width=0.97\linewidth]{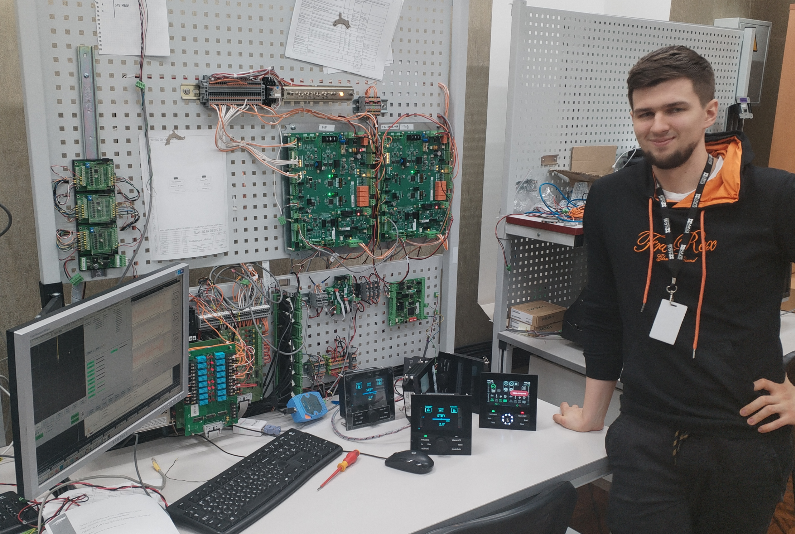}
    \caption{Testbench (HIL) of roll damping system}
    \label{fig:test}
\end{figure}

In this section, we present the results of hardware-in-the-loop (HIL) testing of the control algorithm on a realistic mathematical model of the ship, whose specifications are summarized in the next subsection. 
Vessel performance—both without control and with the roll-damping controller engaged—was evaluated on a roll-damping testbench (Fig.~\ref{fig:test}) comprising a digital-twin simulation computer~\citep{EAmbr2014-DPConf,EAmbr2023-MIT} equipped with analog and digital interfaces that emulate the vessel's onboard equipment and sensors (with measurement noise and faults).
By interfacing the actual roll-damping control hardware with the test bench, we can evaluate the control and identification algorithms under realistic sampling and latency conditions and assess the impact of communication dropouts and other hardware faults. Using the advanced motion models in the digital twin, we evaluate the control algorithms’ performance under realistic noise and disturbance conditions. One can modify the fin electrohydraulic actuators’ nonlinear characteristics (see Fig.~\ref{fig:act}), such as adjusting the opening and closing speeds of the hydraulic distributor’s control valves.

Unless otherwise stated, all experiments employ a polyharmonic wave model with $N=10$ harmonics to approximate the Pierson–Moskowitz spectrum for the relevant sea‐state codes~\citep{Chakrabarti-Handbook,perez2006ship}.

\subsection{Ship Principal Characteristics}\label{subsec.ship}

Key vessel characteristics are summarized in Table~\ref{tab:ship}.
\begin{table}[htb]
\centering
\begin{tabular}{|l|c|}
\hline
   Ship length $L$, m  & 49.9 \\ \hline 
   Ship length $B$, m  & 9.85 \\ \hline
   Ship maximum speed $V$, knt  & 16  \\ \hline
   Fin area $S$, $m^2$  & 3 \\ \hline
   Fin maximal rate $\omega_{f max}$, deg/s  & 35 \\ \hline
   Fin maximal angle $\delta_{f max}$, deg  & 60 \\ \hline
   Actuator time  constant $T_\delta$, s  &  0.14  \\ \hline
   Transversal metacentric height, m  &  0.522  \\ \hline
   Roll natural period at zero speed $T_0=2\pi/\omega_0$, s  &  11  \\
   
   \hline
\end{tabular}
    \caption{Ship particulars}
    \label{tab:ship}
\end{table}



\begin{figure}[htb]
    \centering
    \includegraphics[width=0.7\linewidth]{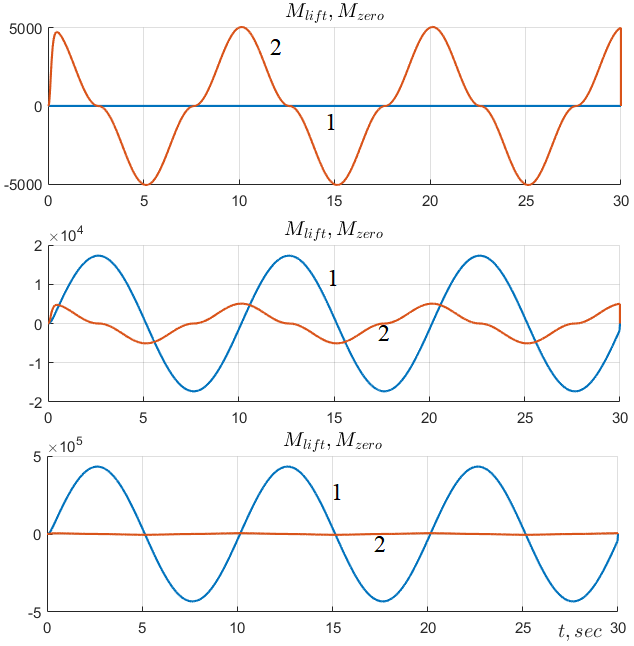}
    \caption{Fin stabilizers control moments for 0, 3, 15 knots (1 -- $M_{\rm lift}$, 2 -- $M_{\rm zero}$) }
    \label{fig:mom}
\end{figure}

The following simulation (Fig.~\ref{fig:mom}) compares the lift‐based and zero-speed drag components of the control moment from \eqref{eq:Mu} at different vessel speeds. Here, with zero disturbance, the control input -- the commanded fin deflection angle -- varies harmonically as
$u(t)=A_u\sin(\omega_u t)$. Because the zero‐speed drag moment $M_{\rm zero}$ depends nonlinearly on the fin’s angular rate $\omega_f=\dot\delta_f$, the resulting $M_{\rm zero}(t)$ is non-harmonic, and its phase is shifted relative to the lift‐induced moment $M_{\rm lift}$, which is proportional to $\delta_f$. The amplitude of $M_{\rm zero}$ remains nearly constant with vessel speed, whereas the amplitude of $M_{\rm lift}$ increases quadratically with speed. Moreover, the absolute amplitudes depend on the control‐signal amplitude ($A_u=20^\circ$ in this test) and its period ($T_u=2\pi/\omega_u=5$s). Standard fin stabilizers rely on lift forces and are thus effective only at sufficiently high inflow speeds ($V_f>5$ kn).


\subsection{Model identification}

The roll‐stabilization system supports open‐loop operation, allowing direct fin angle commands $u(t)$. We perform identification maneuvers in this mode and then determine the model coefficients in the differential equation
\begin{equation}\label{eq:linear1}
  \ddot \theta = - 2 n_\theta  \dot \theta - \omega_0^2\theta  +
    k_1 V_f^2 \delta_f + k_{02} \dot\delta_f |\dot\delta_f|
\end{equation}
using standard least‐squares estimation. The equation~\eqref{eq:linear1} arises from the roll dynamics~\eqref{eq:linear} by decomposing the control moment into the lifting moment (proportional to $V_f^2\delta_f$) and the drag moment (proportional to $\dot\delta_f |\dot\delta_f|$).

Note that the model parameters depend on vessel speed; therefore, identification maneuvers must be conducted across the vessel’s full operational speed range. In the experiment below, we have $V_f\approx 0$, hence, the term $k_1V_f^2\delta_f$ is considered to be negligibly small; we thus only estimate the ship natural frequency  $\omega_0$, dimensionless damping coefficient $\nu_\theta$ and the fin efficiency $k_{02}$.

\begin{figure}[htb]
    \centering
    \includegraphics[width=0.95\linewidth]{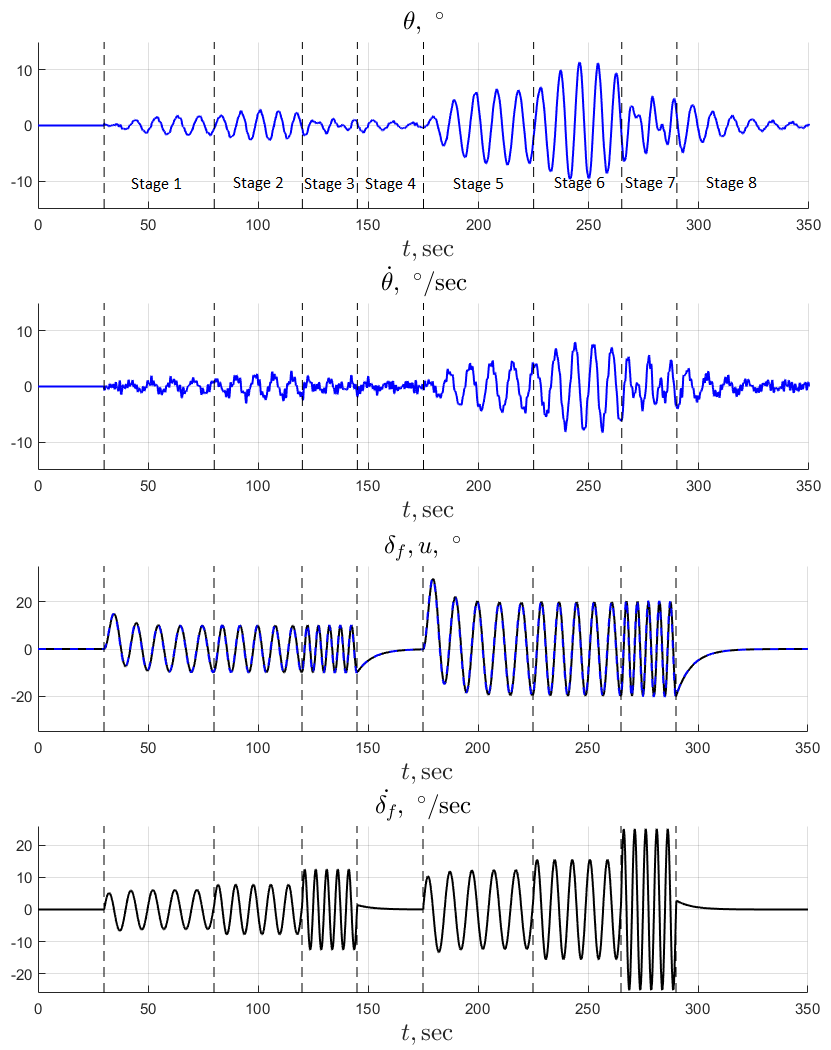}
    \caption{Roll motion in active identification maneuvers}
    \label{fig:ident1}
\end{figure}

We propose a multi-stage calibration maneuver in which, at each stage, the fin command is a harmonic signal  
$u(t) = a_i \sin(\omega_i t)$, with stage-specific amplitudes $a_i$ and frequencies $\omega_i$. Between signals of different amplitude, the command is held at zero for a brief interval. Table~\ref{tab:calibr} and Fig.~\ref{fig:ident1} give an example of parameter values; in this example, the maximum period $T_i = 2\pi/\omega_i$ reaches approximately $0.8\, (2\pi/\omega_0)$. When designing this maneuver, we use an apriori estimation of $\omega_0$ value.

Fig.~\ref{fig:ident1} also shows the rolling dynamics during the calibration manoeuvre.  


For example, the identification results for maneuver (nonlinear model testbench) shown in fig.\ref{fig:ident1} are shown in table\ref{tab:ident}.

\begin{table}[htb]
\centering
  \begin{tabular} {|l| l| l| l|}
   \hline
          & $\omega_0$ & $\nu_\theta$ & $k_{01}$ \\
   \hline
     Identification &  0.729 & 0.060 &  0.1188 \\
   \hline
     Model          &  0.698 & 0.073 &  0.1078 \\
   \hline
  \end{tabular}
  \caption{Identification results}\label{tab:ident}
  \end{table}

\begin{table}[htb]
\centering
  \begin{tabular} {|l| l| l| l|}
   \hline
  Stage i  & $a$  & $T = 2\pi/\omega$  & Duration \\
  \hline
  1        & 10 &   10  &   $5T$ \\  \hline
  2        & 10 &   8  &   $5T$ \\  \hline
  3        & 10 &   5  &   $5T$ \\  \hline
  4        & 0 &   -   &   30 sec \\  \hline
  5        & 20 &   10  &   $5T$ \\  \hline
  6        & 20 &   8  &   $5T$ \\  \hline
  7        & 20 &   5  &   $5T$ \\  \hline
  8        & 0 &   -   &   30 sec \\  \hline
  \end{tabular}
  \caption{Characteristics of the maneuvers}\label{tab:calibr}
\end{table}

\subsection{Controlled Roll Dynamics}

\begin{figure}[htb]
    \centering
    \includegraphics[width=0.7\linewidth]{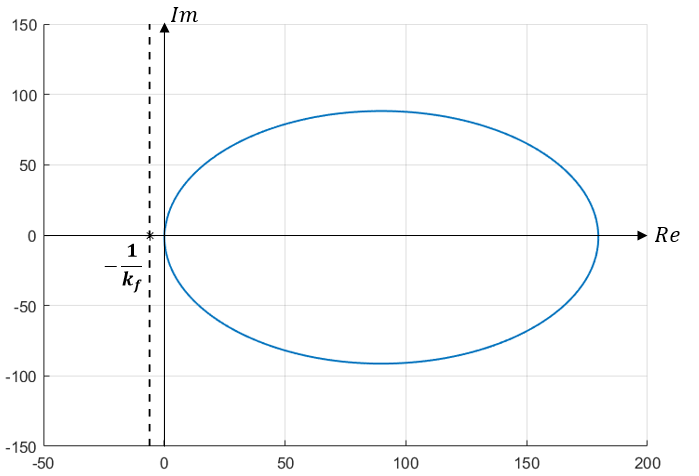}
    \caption{ The Nyquist curve $\tilde W(j\omega)$}
    \label{fig:hodo}
\end{figure}

Finally, we demonstrate the roll response under the controller defined in~\eqref{eq:controller1} using the tuned gains $k_p = 2.3$, $k_d = 15.1$, and $c = 0.14$. Although the Nyquist plot comes sufficiently close to the critical point $-1/k_f$ (Fig.~\ref{fig:hodo}), the frequency-domain condition~\eqref{eq:freq} remains satisfied.



\begin{figure}
    \centering
    \includegraphics[width=0.85\linewidth]{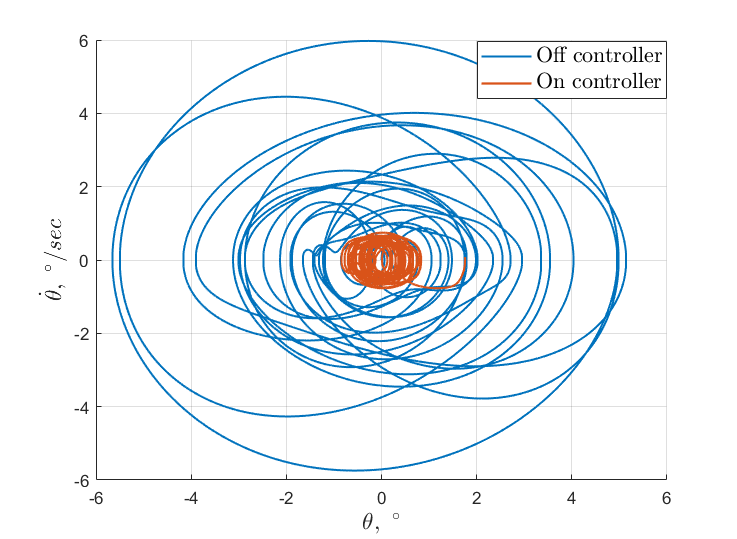}
    \caption{ Phase portrait for wave with sea state code 5   ($H_s=2.2m$, $T_z=5.4s$)
    }
    \label{fig:controller1}
\end{figure}

To illustrate the controller’s effectiveness, Fig.~\ref{fig:controller1} compares the roll-model phase portraits with the controller off and on. Upon activation, the trajectory converges more rapidly toward the phase-plane origin. In these experiments, the vessel maintains an approximately fixed position (its forward speed is neglected) and constant heading, with waves approaching at a $90^\circ$ angle from starboard.

%





\section*{Conclusion}

This paper presented control and identification strategies for a zero-speed fin-based roll stabilization system developed by Navis JSC. We addressed key challenges associated with the nonlinear nature of the drag-induced torque and the physical constraints of the hydraulic fin actuators. A two-stage control scheme was introduced to decouple the generation of desired angular rate commands from the actuator's angle-based input, ensuring compatibility with hardware limitations. Furthermore, we highlighted that traditional linear control methods, while widely used, require careful stability analysis when applied to systems with nonlinear rate–torque characteristics. Using the frequency-domain contraction criterion, we established conditions for the incremental stability of the closed-loop system. Simulation studies, based on a realistic ship model, confirmed the performance and practicality of the proposed control approach for roll damping in low-speed and zero-speed scenarios. 

Note that this work focuses on low-speed roll damping, which represents the primary limitation. A complete solution would involve tuning both low-speed and high-speed controllers, along with implementing a bumpless switching logic between them. This, together with a systematic procedure for control gain selection and experimental validation under varying sea states and vessel loading conditions, will be the subject of future research.
In this work, we did not address the problem of model identification under varying environmental conditions, as our identification experiments were limited to the models used for hardware-in-the-loop (HIL) testing.


\bibliography{ref}
\end{document}